\documentclass[prd]{revtex4}
\usepackage{amssymb}
\usepackage{bm}
\newcommand{\cc}{\cite}

\newcommand{\be}{\begin{equation}}
\newcommand{\ee}{\end{equation}}
 
\def\ort{\hbox{$\cal T$}}

\def\pd{\partial}

\def\L{\Lambda}

\def\<{\langle}
\def\>{\rangle}

\def\ch{\cosh}

\def\a{\alpha}
\def\b{\beta}
\def\g{\gamma}  \def\G{\Gamma}
\def\d{\delta}  
\def\l{\lambda}   \def\L{\Lambda}
\def\s{\sigma}
\def\r{\rho}  
\def\x{\xi}
\def\c{\chi}
\def\m{\mu}
\def\n{\nu}
\def\t{\tau}

\def\({\left(}
\def\[{\left[}
\def\){\right)}
\def\]{\right]}
\def\coth{\hbox{coth}}

\def\pd{\partial}

\def\pa{{\cal P}}

\begin{document}
\title{\bf High energy behavior of
quark elastic form factors \protect \\ within the Wilson
integral approach: \protect \\ perturbative and nonperturbative
contributions\footnote{Talk given at the XVII International
Workshop on High Energy Physics and Quantum Field Theory (QFTHEP
2003), Samara-Saratov, Russia, 4-11 Sept 2003; to be published
in the Proceedings.}}
\author{Igor O. Cherednikov}
\email{igorch@thsun1.jinr.ru, igor.cherednikov@jinr.ru}
\affiliation{\sl
Joint Institute for Nuclear Research \\
RU-141980 BLTP JINR, Dubna, Russia}
\date{\today}
\begin{abstract}
The Wilson contour integral approach is applied to resum the
soft gluon radiative correctins to the quark form factors in the
Sudakov regime. The one-loop order results for the quark-photon
(color singlet form factor) and quark-gluon (color non-singlet
form factor) vertices are presented. The explicit expressions
for the vacuum averaged contour integrals in $g^2$ accuracy are
derived for an arbitrary gauge field.  The corresponding
one-loop cusp anomalous dimensions are found in the case of
perturbative gluon field in arbitrary covariant gauge. It is
shown that the gauge dependence drops out from the leading high
energy behavior.
\end{abstract}
\maketitle

In this report, the brief summary of the recent study (within
the framework of the world-line contour integrals) of the
perturbative and nonperturbative contributions to the soft gluon
radiative effects for the quark form factors at large
transferred momenta is given.

In sector of the strong interactions of the Standard Model, the
elastic form factors of quarks are the most elementary entities
exhibiting the large resumed logarithmic corrections at large
transferred momentum. These quantities enter into the
quark-photon and quark-gluon vertices in the calculations of
various QCD processes at the partonic level, and are under
active investigation nowadays \cc{HARD, FEN, SIM, JP}.

Here we apply the powerful Wilson contour integral techniques
\cc{NACH, KRON, STEF} to perform the Sudakov resummation of soft
gluon radiative corrections to the {\it quark-vector boson}
vertices with large transferred momentum. The world-line
formulation of a quantum theory is actively developed not only
due to the wide range of applications in the perturbative QFT,
but also from the point of view of the various string theories
(for a recent review, see \cc{STR} and Refs. therein). The
attractive feature of this approach is that it does not refer
directly to the standard perturbative techniques allowing one to
avoid the explicit diagrammatic calculations which use to be
very involved in non-Abelian gauge theories. Therefore, this
method is equally suitable for the perturbative as well as
nonperturbative calculations \cc{ST}. In this report, we
demonstrate how the Wilson integrals formalism can be applied
for arbitrary gauge fields in $n$-dimensional space-time, in a
particular case of the on-shell quark form factor.

Among a number of independent form factors entering the {\it
quark-vector boson} vertex, the only one contains IR
singularities and is not power-suppressed in the Sudakov
kinematics (see, {\it e.g.}, \cc{QGV, CH} and Refs. therein).
This form factor can be defined via the amplitude of the elastic
quark scattering in an external gauge field---the quark on a
mass shell comes from infinity, emits the hard vector boson at
the origin, and goes away to infinity: \be u_i(p)
\[{\cal M}_\m \]^a_{ij} v_j(p')= F\[(p-p')^2 ; \x
\] \bar u_i(p) t^a_{ij}\g_\m v_j(p') \ \ , \label{eq:ampl}
\ee where $p,p'$ are the momenta of the in-coming and out-going
quarks, and $\x$ is the covariant gauge-fixing  parameter. In
this formulae, we write down the vector boson's and quark's
color indices $a, (i,j)$ for generality, assuming that the
external vector boson is a colored gluon \cc{CH}. The case of a
photon is trivially restored by means of the replacement of
$t^a_{ij}$ with unity matrix. The Sudakov kinematics is
determined by the small masses of the quarks and large squared
transferred momentum: \be m^2 = p^2 = {p'}^2 \ \ , \ \ (pp') =
m^2 \ch \c \ , \ (pp') >> m^2 \ee

Following the Refs. \cc{KRON, STEF}, we express the IR sensitive
contribution of the resumed soft gluon radiative corrections to
the form factor as the vacuum average of  two path-ordered
exponentials \be t_{ij}^a \ F \[Q^2; \x\] = \Big<0\Big|\ort
\Big\{ W_{ii'} t^a_{i'j'} W_{j'j} \Big\}\Big|0\Big>  \ ,
\label{eq:def1} \ee where \be W_{ii'} = \pa \exp \[ig \ t^\a
v_\m \int_{-\infty}^0 \! d\s \ A^\a_\m (v \s) \] \Bigg|_{ii'} \
,  \  W_{j'j} = \pa \exp \[ig \ t^\b v'_\m \int_{0}^\infty \!
d\s \ A^\b_\m (v' \s) \]\Bigg|_{j'j} \ . \ee The quark
trajectories can be parameterized as: \be {\bm In:} \ \ x_\m =
v_\m \ \t \ \ , \ \ \t \in [-\infty, 0] \ \ , \ \ v_\m = p_\m /
m \ , \ {\bm Out:} \ \ y_\n = v'_\n \ \s \ \ , \ \ \s \in [0,
+\infty] \ \ , \ \ v'_\n = p'_\n / m \ . \ee For calculations to
the $O(g^2)$ accuracy, it is convenient to present the gauge
field propagator $D_{\m\n}(z)$: \be \Big<0\Big|\ort A^\a_\m (x)
A^\b_\n (y)\Big|0\Big> =
 {\cal D}_{\m\n}^{\a\b}(x-y) = \d^{\a\b} D_{\m\n}(x-y)\ee
in the form \cc{OUR}: \be D_{\m\n}(z) = g_{\m\n} \pd_\r \pd^\r
D_1(z^2) - \pd_\m\pd_\n D_2(z^2) \ . \label{st1} \ee First, we
present the vacuum averaged Wilson integral Eq. (\ref{eq:def1})
in $n$-dimensional space-time, for an arbitrary gauge field
which can be of any origin, for instance, it may be
nonperturbative. The leading order $ \sim g^2$ terms stem from
the expressions: \be W^{(1)}_{LO} = - {g^2 \over 2}\ t^a_{ij} \
C_F \ v_\m v_{\m'} \ \int_{0}^\infty \! d\s \int_0^\infty \!
d\s' \  D_{\m\m'}
\[v(\s-\s')\] \ , \label{eq:qdep0} \ee \be W^{(2)}_{LO} = -
{g^2 \over 2}\ t^a_{ij} \ C_F \ v'_\m v'_{\m'} \
\int_{-\infty}^0 \! d\s \int_{-\infty}^0 \! d\s' \ D_{\m\m'}
\[v'(\s-\s')\] \ , \ee and \be W^{(12)}_{LO} = - {g^2 \over 2}\
G_F \  t^a_{ij}\ v_\m v'_\n\ \int_{0}^\infty \! d\t
\int_0^\infty \! d\s \ D_{\m\n} (v\t+v'\s) \ . \label{eq:qdep1}
\ee The general result in $n$ dimensions and arbitrary covariant
gauge, with the gauge field two-point correlator expressed as
Eq. (\ref{st1}), reads: \be W^{(1)} = W^{(2)} = - t^a_{ij}\
\frac{g^2}{2} C_F \ \[(n-2) D_1(-b_{\perp}^2) + 2 b^2_\perp
D'_1(-b_{\perp}^2) + D_2 (-b_{\perp}^2) \] \ , \label{gen1}\ee
\be W^{(12)}(\c) = t^a_{ij}\  g^2 G_F
\[\c\coth \c \((n-2) D_1(-b_{\perp}^2) + 2 b_{\perp}^2 D'_1(-b_{\perp}^2)
\) + D_2(-b_{\perp}^2)\] , \label{gen2}\ee up to $O(g^4)$ order
terms. Here $C_F = \(N_c^2 - 1\)/2N_c$ is the quadratic Casimir
operator in the fundamental representation, and $G_F$ is the
color factor which is \be G_F^S = C_F \ee for the external
photon field (color singlet form factor) and \be G_F^{NS} =  C_F
- \frac{C_A}{2} \ , \  C_A = N_c \ , \ee for the gluon probe
(color non-singlet form factor). The space-like transversal
vector $\vec b_\perp$ is introduced in order to regulate the UV
divergence.

Performing the standard renormalization procedure within the
${\overline{MS}}$ scheme, described in detail in Refs. \cc{WREN,
MMP, KRON}, one finds for the {\it perturbative gluon field}
(with $\l^2$ being the IR cutoff) \be W_{LO}^{(1)}(\a_s,
\m^2/\l^2; \x) = W_{LO}^{(2)}(\a_s, \m^2/\l^2; \x) = t^a_{ij}
\frac{\a_s}{4\pi} C_F \( 1 - \frac{\x}{2} \)\ \ln
\frac{\m^2}{\l^2}  \ , \label{w1} \ee and \be W_{LO}^{(12)}
(\a_s, \c, \m^2/\l^2; \x) = - t^a_{ij} \frac{\a_s}{2\pi} G_F \
\[ \c \coth \c - \frac{\x}{2}
\] \ln \frac{\m^2}{\l^2} \ . \label{w12} \ee Here the
UV-normalization point is taken to be $\m^2 = 4 \vec b^{-2}$.
The total one-loop contribution to the form factor is given by
the sum \be F_{LO} \[Q^2; \x\] = W_0 + 2 W_{LO}^{(1)} \(\a_s,
\frac{\m^2}{\l^2}\) + W_{LO}^{(12)}\(\a_s, \frac{\m^2}{\l^2}, \c
\) + O(\a_s^2)\ . \ee

The high-energy asymptotic behavior of the form factor is
determined by the (in general, gauge-dependent) cusp anomalous
dimension which is derived from the renormalization of the
Wilson integral (\ref{eq:def1}) \cc{KRON}: \be \(\m^2
\frac{\pd}{\pd \m^2} + \b(\a_s)\frac{\pd}{\pd \a_s} + \d (\a_s,
\x) \x \frac{\pd}{\pd \x} \) \ln F (\c) = - \frac{1}{2}
\G_{cusp}
\[\a_s(\m^2); \c \] \ . \label{ad} \ee This anomalous dimension
reads in our case: \be \G_{cusp}
\[\a_s(\m^2); \c\] = \frac{\a_s}{\pi} \[ G_F
\c \coth \c + \frac{C_F - G_F }{2} \x - C_F  \] + O(\a_s^2)\ .
\ee It is easy to see that for the color singlet case, the gauge
dependent term is cancelled, as it should be for the gauge
invariant quantity.

Taking into account that the cusp anomalous dimension is linear
in $\ln Q^2$ at large $\c$ \cc{KRON}: \be \G_{cusp} (C_\c; \a_s)
= \ln q^2 \ \G_{cusp} (\a_s) + O(\ln^0 q^2)\ , \ee one obtains
the leading (double-logarithmic) behavior of the quark form
factor from the corresponding evolution equation:
$$
F\[Q^2\] = \exp\(-G_F\int_{\l^2}^{Q^2}\! \frac{d\m}{2\m} \
\ln\frac{Q^2}{\m}\ \frac{\a_s(\m)}{\pi}  + \hbox{NLO terms} \) F
\[\a_s(\l^2)\] = $$ \be = \exp\[- \frac{2G_F}{\b_0} \ln q^2
\ln\ln q^2 + O(\ln q^2) + \hbox{NLO terms}\] F
\[\a_s(\l^2)\] \ ,  \label{evolfin} \ee where
$q^2 = Q^2/\L_{QCD}^2$ is dimensionless variable. In any case,
the dependence from the gauge-fixing parameter $\x$ drops out of
the leading logarithmic expression for $\G_{cusp}$, and yields
no influence on the main asymptotics. Thus we find that the cusp
anomalous dimension of non-singlet quark form factor is
negative: $$ \G_{cusp}^{NS}
\[\a_s(\m^2); \c \] = $$ \be = \frac{\a_s}{\pi} \[ \(C_F - \frac{C_A}{2}\)
\c \coth \c + \frac{C_A}{4} \x - C_F  \] + O(\a_s^2) = -
\frac{\a_s}{\pi} \[\frac{1}{2N_c} \c \coth \c - \frac{N_c}{4} \x
- \frac{N_c^2 -1}{2N_c} \ , \] \ee that corresponds to the
enhancement of the resumed Sudakov logarithms at large $Q^2$,
while in the singlet case it is positive: $$ \G_{cusp}^{S}
\[\a_s(\m^2); \c \] = $$ \be = \frac{\a_s}{\pi} \[ C_F
\c \coth \c +  \(C_F - C_F \)\x - C_F  \] + O(\a_s^2) =
\frac{\a_s}{\pi} \frac{N_c^2 -1}{2N_c} \[  \c \coth \c -1 \] +
O(\a_s^2)\ , \ee that yields the well known Sudakov suppression.

The formulas (\ref{gen1}, \ref{gen2}) can be applied directly to
compute the leading contributions to the Wilson vacuum average
for an arbitrary $SU(N_c)$ gauge field in any covariant gauge,
in $n$-dimensional space-time. For instance, the nonperturbative
instanton-induced corrections (within the framework of the
Instanton Liquid Model (ILM) of QCD vacuum \cc{ILM}), as well as
IR-renormalon effects, are evaluated within this approach in
Refs. \cc{OUR}.

The present research is partially supported by RFBR (Grant Nos.
03-02-17291, 02-02-16194), Russian Federation President's Grant
1450-2003-2, and INTAS (Grant No. 00-00-366).

\end{document}